\documentclass[%
 reprint,
superscriptaddress,
showpacs,
 amsmath,amssymb,
 aps,
 prl, 
]{revtex4-2}

\usepackage{graphicx}
\usepackage{dcolumn}
\usepackage{bm}
\usepackage{xcolor}
\usepackage[caption=false]{subfig}

\begin{document}

\preprint{APS/123-QED}

\title{Search for Neutrino-Induced Neutral Current $\Delta$ Radiative Decay in MicroBooNE and a First Test of the MiniBooNE Low Energy Excess Under a Single-Photon Hypothesis}


\newcommand{\Bern}{Universit{\"a}t Bern, Bern CH-3012, Switzerland}
\newcommand{\BNL}{Brookhaven National Laboratory (BNL), Upton, NY, 11973, USA}
\newcommand{\UCSB}{University of California, Santa Barbara, CA, 93106, USA}
\newcommand{\Cambridge}{University of Cambridge, Cambridge CB3 0HE, United Kingdom}
\newcommand{\CIEMAT}{Centro de Investigaciones Energ\'{e}ticas, Medioambientales y Tecnol\'{o}gicas (CIEMAT), Madrid E-28040, Spain}
\newcommand{\Chicago}{University of Chicago, Chicago, IL, 60637, USA}
\newcommand{\Cincinnati}{University of Cincinnati, Cincinnati, OH, 45221, USA}
\newcommand{\CSU}{Colorado State University, Fort Collins, CO, 80523, USA}
\newcommand{\Columbia}{Columbia University, New York, NY, 10027, USA}
\newcommand{\Edinburgh}{University of Edinburgh, Edinburgh EH9 3FD, United Kingdom}
\newcommand{\FNAL}{Fermi National Accelerator Laboratory (FNAL), Batavia, IL 60510, USA}
\newcommand{\Granada}{Universidad de Granada, Granada E-18071, Spain}
\newcommand{\Harvard}{Harvard University, Cambridge, MA 02138, USA}
\newcommand{\IIT}{Illinois Institute of Technology (IIT), Chicago, IL 60616, USA}
\newcommand{\KSU}{Kansas State University (KSU), Manhattan, KS, 66506, USA}
\newcommand{\Lancaster}{Lancaster University, Lancaster LA1 4YW, United Kingdom}
\newcommand{\LANL}{Los Alamos National Laboratory (LANL), Los Alamos, NM, 87545, USA}
\newcommand{\Manchester}{The University of Manchester, Manchester M13 9PL, United Kingdom}
\newcommand{\MIT}{Massachusetts Institute of Technology (MIT), Cambridge, MA, 02139, USA}
\newcommand{\Michigan}{University of Michigan, Ann Arbor, MI, 48109, USA}
\newcommand{\Minnesota}{University of Minnesota, Minneapolis, MN, 55455, USA}
\newcommand{\NMSU}{New Mexico State University (NMSU), Las Cruces, NM, 88003, USA}
\newcommand{\Oxford}{University of Oxford, Oxford OX1 3RH, United Kingdom}
\newcommand{\Pitt}{University of Pittsburgh, Pittsburgh, PA, 15260, USA}
\newcommand{\Rutgers}{Rutgers University, Piscataway, NJ, 08854, USA}
\newcommand{\SLAC}{SLAC National Accelerator Laboratory, Menlo Park, CA, 94025, USA}
\newcommand{\SDSMT}{South Dakota School of Mines and Technology (SDSMT), Rapid City, SD, 57701, USA}
\newcommand{\Maine}{University of Southern Maine, Portland, ME, 04104, USA}
\newcommand{\Syracuse}{Syracuse University, Syracuse, NY, 13244, USA}
\newcommand{\TelAviv}{Tel Aviv University, Tel Aviv, Israel, 69978}
\newcommand{\Tennessee}{University of Tennessee, Knoxville, TN, 37996, USA}
\newcommand{\UTA}{University of Texas, Arlington, TX, 76019, USA}
\newcommand{\Tufts}{Tufts University, Medford, MA, 02155, USA}
\newcommand{\VTech}{Center for Neutrino Physics, Virginia Tech, Blacksburg, VA, 24061, USA}
\newcommand{\Warwick}{University of Warwick, Coventry CV4 7AL, United Kingdom}
\newcommand{\Yale}{Wright Laboratory, Department of Physics, Yale University, New Haven, CT, 06520, USA}

\affiliation{\Bern}
\affiliation{\BNL}
\affiliation{\UCSB}
\affiliation{\Cambridge}
\affiliation{\CIEMAT}
\affiliation{\Chicago}
\affiliation{\Cincinnati}
\affiliation{\CSU}
\affiliation{\Columbia}
\affiliation{\Edinburgh}
\affiliation{\FNAL}
\affiliation{\Granada}
\affiliation{\Harvard}
\affiliation{\IIT}
\affiliation{\KSU}
\affiliation{\Lancaster}
\affiliation{\LANL}
\affiliation{\Manchester}
\affiliation{\MIT}
\affiliation{\Michigan}
\affiliation{\Minnesota}
\affiliation{\NMSU}
\affiliation{\Oxford}
\affiliation{\Pitt}
\affiliation{\Rutgers}
\affiliation{\SLAC}
\affiliation{\SDSMT}
\affiliation{\Maine}
\affiliation{\Syracuse}
\affiliation{\TelAviv}
\affiliation{\Tennessee}
\affiliation{\UTA}
\affiliation{\Tufts}
\affiliation{\VTech}
\affiliation{\Warwick}
\affiliation{\Yale}

\author{P.~Abratenko} \affiliation{\Tufts} 
\author{R.~An} \affiliation{\IIT}
\author{J.~Anthony} \affiliation{\Cambridge}
\author{L.~Arellano} \affiliation{\Manchester}
\author{J.~Asaadi} \affiliation{\UTA}
\author{A.~Ashkenazi}\affiliation{\TelAviv}
\author{S.~Balasubramanian}\affiliation{\FNAL}
\author{B.~Baller} \affiliation{\FNAL}
\author{C.~Barnes} \affiliation{\Michigan}
\author{G.~Barr} \affiliation{\Oxford}
% ...

\collaboration{The MicroBooNE Collaboration}
\thanks{microboone\_info@fnal.gov}\noaffiliation

\date{\today}

\begin{abstract}
We report results from a search for neutrino-induced neutral current (NC) resonant $\Delta$(1232) baryon production followed by $\Delta$ radiative decay, with a $\langle0.8\rangle$~GeV neutrino beam. Data corresponding to MicroBooNE's first three years of operations (6.80$\times$10$^{20}$ protons on target) are used to select single-photon events with one or zero protons and without charged leptons in the final state ($1\gamma1p$ and $1\gamma0p$, respectively). The background is constrained via an \textit{in-situ} high-purity measurement of NC $\pi^0$ events, made possible via dedicated $2\gamma1p$ and $2\gamma0p$ selections. A total of 16 and 153 events are observed for the $1\gamma1p$ and $1\gamma0p$ selections, respectively, compared to a constrained background prediction of $20.5 \pm 3.65 \text{(sys.)} $ and  $145.1 \pm 13.8 \text{(sys.)} $  events. The data lead to a bound on an anomalous enhancement of the normalization of NC $\Delta$ radiative decay of less than $2.3$ times the predicted nominal rate for this process at the 90\% confidence level (CL). The measurement disfavors a candidate photon interpretation of the MiniBooNE low-energy excess as a factor of $3.18$ times the nominal NC $\Delta$ radiative decay rate at the 94.8\% CL, in favor of the nominal prediction, and represents a  greater than $50$-fold improvement over the world's best limit on single-photon production in NC interactions in the sub-GeV neutrino energy range.

\end{abstract}

\maketitle

For over two decades, the anomalous signals consisting of MiniBooNE's low-energy excess (LEE) \cite{MiniBooNE:2008yuf,MiniBooNE:2018esg,Aguilar_Arevalo_2021} and the prior LSND \cite{LSND:1996vlr} $\overline{\nu}_e$ appearance results have been at the forefront of neutrino physics. Each has been interpreted as evidence for new types of neutrinos or other physics beyond the Standard Model (SM). The existence of new particles would be the first evidence for a new paradigm of physics associated with the neutrino sector since the discovery of neutrinos mass via their observed oscillations, and would have profound ramifications for all particle physics, astrophysics, and cosmology. At the heart of this puzzle of anomalies in need of interpretation is the fact MiniBooNE could not differentiate neutrino interactions producing an electron (such as from $\nu_e$ appearance due to light sterile neutrinos) from those with a single photon in the final state. Thus, both types of interactions must be examined independently as a source of the LEE.

Neutrino-induced neutral current (NC) production of the $\Delta$(1232) baryon resonance with subsequent $\Delta$ radiative decay is predicted to be the dominant source of single photons in neutrino-argon scattering below 1~GeV \cite{Wang_2014}. Although $\Delta$ radiative decay is predicted in the SM, and measurements of photoproduction \cite{Blanpied:1997zz} and virtual compton scattering \cite{Sparveris:2008jx} are well described by theory, this process has never been directly observed in neutrino scattering. Previous searches have been performed by the T2K~\cite{Abe_2019} and NOMAD~\cite{Kullenberg_2012} experiments with average incident neutrino energies, $E_\nu$, of 0.85 and 25~GeV, respectively, resulting in leading limits on this process. Although on a different target, T2K's result is closest in $E_\nu$ to that of the MiniBooNE beam. However, the 90\% confidence level (CL) limit is $\sim100$ times the theoretically predicted rate of NC $\Delta$ radiative decay. 

In this letter, we present the world's most sensitive search for NC $\Delta\rightarrow N \gamma$, where $N=p,n$, using neutrino-argon scattering data collected by the MicroBooNE detector \cite{Acciarri_2017}. MicroBooNE is an 85 metric ton active volume liquid argon time projection chamber (LArTPC) situated at a similar baseline in the same muon neutrino dominated Booster Neutrino Beam (BNB) at Fermilab~\cite{PhysRevD.79.072002} as MiniBooNE, with $\langle E_\nu\rangle = 0.8$~GeV. The measurement makes use of data corresponding to a BNB exposure of $6.80\times10^{20}$~protons on target (POT), collected during 2016-2018. LArTPC technology allows MicroBooNE to distinguish electromagnetic showers originating from electrons or photons based on ionization energy deposition ($dE/dx$) at the start of the shower, and the non-zero conversion distance of the photon relative to the interaction vertex. 

This search represents a first for this process with argon as the neutrino target, and also constitutes the first test of the MiniBooNE LEE under a single-photon interpretation. In a fit to the radial distribution of the MiniBooNE data with statistical errors only, an enhancement of NC $\Delta\rightarrow N \gamma$ (as predicted by the NUANCE \cite{nuance-miniboone} neutrino event generator on CH$_{2}$) by a normalization factor of $x_\text{MB}=3.18$ (quoted with no uncertainty) was found to provide the best fit for the observed LEE \cite{Aguilar_Arevalo_2021}. We perform an explicitly model-dependent test of this interpretation, cast as a factor of 3.18 enhancement to the predicted NC $\Delta\rightarrow N \gamma$ rate in MicroBooNE, under a two-hypothesis $\Delta \chi^2$ test between the enhanced rate and the nominal NC $\Delta\rightarrow N\gamma$ prediction. 

MicroBooNE uses a custom tune \cite{MicroBooNE:2021ccs} of the \textsc{genie} neutrino event generator v3.0.6 \cite{Andreopoulos:2009rq,genie_v3} to simulate neutrino-argon interactions. At BNB energies, the dominant source of single-photon production with no charged leptons or pions in the final state is NC $\Delta (1232) \rightarrow N \gamma$. This process is included in the MicroBooNE nominal prediction exactly as modeled in \textsc{genie}. Heavier resonances and non-resonant processes, including coherent single-photon production \cite{coherent}, are not currently included in the simulation, but are each estimated to contribute at the 10\% level or less. Both these processes would produce slightly higher-energy photons than the $\Delta$(1232) resonance, and a more forward (in the direction of the neutrino beam) photon in the case of coherent production. Although such events may be selected by this analysis, we do not explicitly quantify their selection efficiency and in this letter we focus on the dominant NC $\Delta (1232) \rightarrow N \gamma$ process.

The MicroBooNE NC $\Delta\rightarrow N \gamma$ search exclusively targets events with a single, photon-like electromagnetic shower and either no other visible activity or one visible final-state proton. These are referred to as $1\gamma0p$ and $1\gamma1p$ events and primarily probe $\Delta\rightarrow n\gamma$ and $\Delta\rightarrow p\gamma$ decays, respectively. 
The analysis selects and simultaneously fits $1\gamma1p$ and $1\gamma0p$ data-to-Monte Carlo (MC) simulated distributions together with two additional, mutually exclusive but highly correlated event samples: $2\gamma1p$, and $2\gamma0p$. The signal, defined as all true NC $\Delta\rightarrow N\gamma$ events whose true interaction vertex is inside the active TPC, contributes predominantly to the $1\gamma$ event samples. The high-statistics $2\gamma$ samples are enhanced in NC $\Delta\rightarrow N\pi^0$ production, which is the dominant source of mis-identified background to the $1\gamma1p$ and $1\gamma0p$ event samples. 

Reconstruction of all four event samples makes use of the Pandora framework \cite{ub_pandora}. 
Reconstructed ionization charge hits are clustered and matched across three 2D projected views of the MicroBooNE active TPC volume into 3D reconstructed objects. These are then classified as tracks or showers based on a multivariate classifier score and aggregated into candidate neutrino interactions. The topological selection of interactions with exactly one shower and zero or one tracks represents the basis of the $1\gamma$ selections. Subsequently, pre-selection requires that the reconstructed vertex, shower-start point and track (as applicable) are all fully contained within the detector fiducial volume. A minimum {energy} requirement is imposed on the shower,  ensuring good reconstruction performance, and a maximum track length requirement is imposed on the track, rejecting obvious muon backgrounds. 
Tracks are also required to have a high $dE/dx$ consistent with that of a proton. Finally, an opening angle requirement between the track and shower directions is applied to eliminate co-linear events where the start of a shower can be mis-reconstructed as a track.

The pre-selected events are fed into a set of boosted decision trees (BDTs), each designed to reject a distinct background and select NC $\Delta \rightarrow N\gamma$ events. The gradient boosting algorithm XGBoost \cite{Chen:2016} is used to train the BDTs. A cosmic BDT rejects cosmogenic backgrounds and is trained on cosmic ray data events collected when no neutrino beam was present. Track calorimetry is used to reject cosmic muons, with track and shower directionality-based variables proving powerful discriminators. A NC $\pi^0$ BDT compares the relationship of the reconstructed shower and track to those expected from $\pi^{0}$ decay kinematics to separate true single-photon events from those containing a $\pi^{0}$ decay where a second photon is not reconstructed. A charged current (CC) $\nu_e$ BDT targets the intrinsic $\nu_e$ background events. Here, the photon conversion distance and shower calorimetry play important roles. A fourth BDT is designed to veto events in which a second shower from a $\pi^0$ decay deposits some charge, but fails 3D shower reconstruction. Such events can result in 2D charge hits near the neutrino interaction that are not associated with a 3D object. A plane-by-plane clustering algorithm,  DBSCAN \cite{dbscan}, is used to group these unassociated hits, and properties including direction, shape and energy of the cluster are used to determine consistency with a second shower from a $\pi^{0}$ decay. A final CC $\nu_\mu$-focused BDT removes any remaining backgrounds, primarily targeting the muon track through track calorimetry variables.  

The $1\gamma1p$ selection uses all five BDTs. The absence of a track in the $1\gamma0p$ sample means that the $1\gamma0p$ selection cannot use these BDTs identically, as it is limited to only shower variables. As such, it uses variations of the cosmic and NC$\pi^0$ BDTs, and a third BDT merging the functionality of the CC $\nu_e$ and CC $\nu_\mu$-focused BDTs, targeting all remaining backgrounds. All BDTs are trained explicitly to select well-reconstructed NC $\Delta \rightarrow N \gamma$ events. While model-dependent, this leverages the kinematics and correlations between the track and shower associated with $\Delta(1232)$ resonance decay, particularly for the $1\gamma1p$ selection. The BDTs for the $1\gamma1p$ and $1\gamma0p$ selections are trained and optimized for each selection independently, through a scan over grids of their BDT classifier scores. The optimized BDT classifier score cuts correspond to the highest statistical significance of the NC $\Delta\rightarrow N\gamma$ signal over background in each sample. The topological, pre-selection, BDT selection, and combined signal efficiencies are summarized in Table~\ref{tab:efficiencies}.

\begin{table}[t!]
    \begin{tabular}{lrrr}
    \hline\hline
        Selection Stage &  $1\gamma1p$ eff. & $1\gamma0p$ eff. \\ \hline
        Topological & 19.4\% & 13.5\%  \\
        Pre-selection & 63.9\% & 98.4 \% \\
        BDT Selection & 32.1\% & 39.8\%  \\ \hline
        Combined & 3.99\% & 5.29\%  \\ \hline\hline
    \end{tabular}
    \caption{Signal efficiencies for the $1\gamma1p$ and $1\gamma0p$ selections. The topological and combined efficiencies  are evaluated relative to all true NC $\Delta\rightarrow N\gamma$ events inside the active TPC in the simulation (124.1 events expected for $6.80 \times 10^{20}$ POT). The pre-selection and BDT selection efficiencies are evaluated relative to their respective preceding selection stage. }
    \label{tab:efficiencies}
\end{table}

\begin{figure}[t]
    \centering
    \subfloat[]{\includegraphics[trim={0 0 0 0},clip,width=\linewidth]{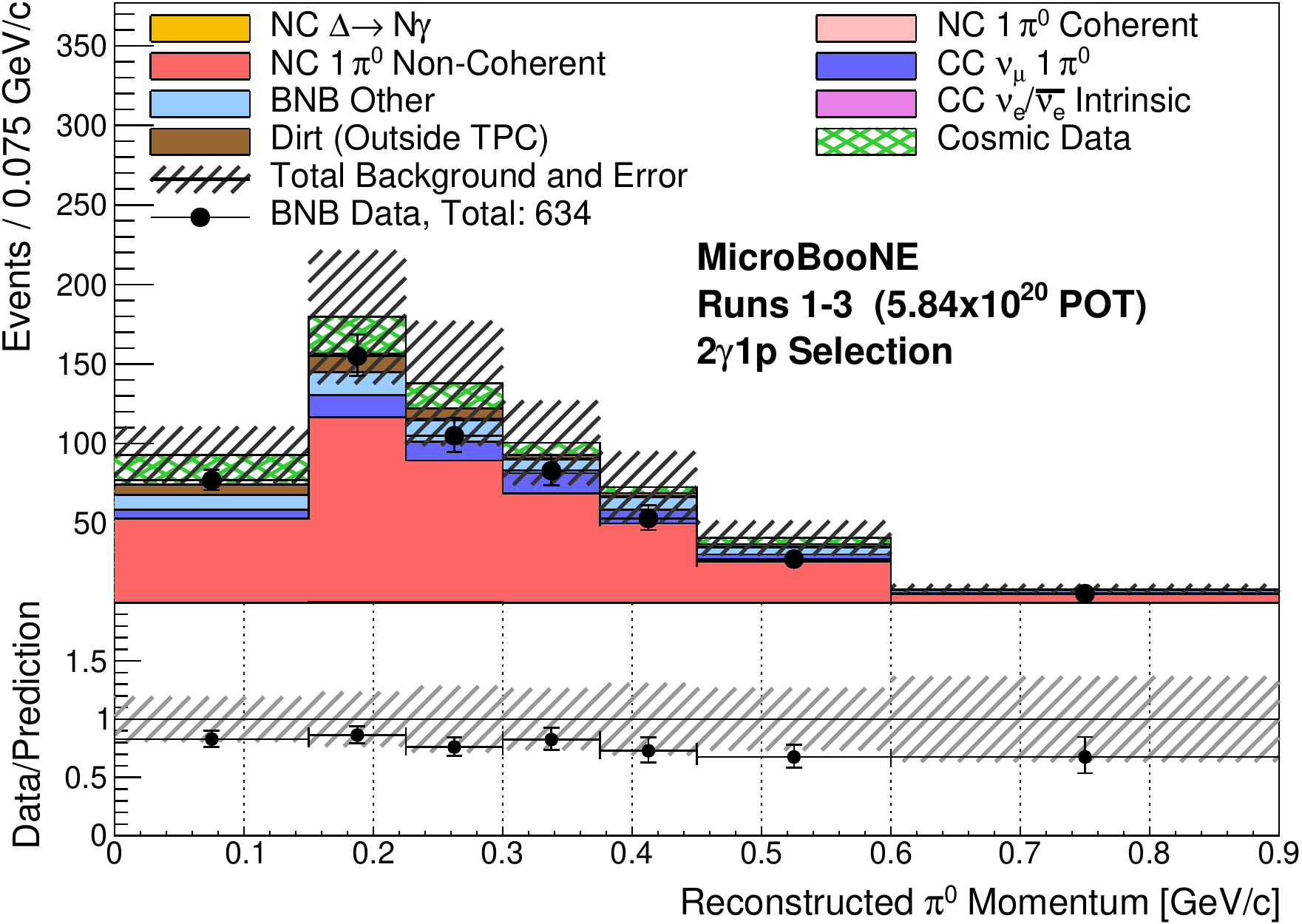}}\\
    \subfloat[]{\includegraphics[trim={0 0 0 0},clip,width=\linewidth]{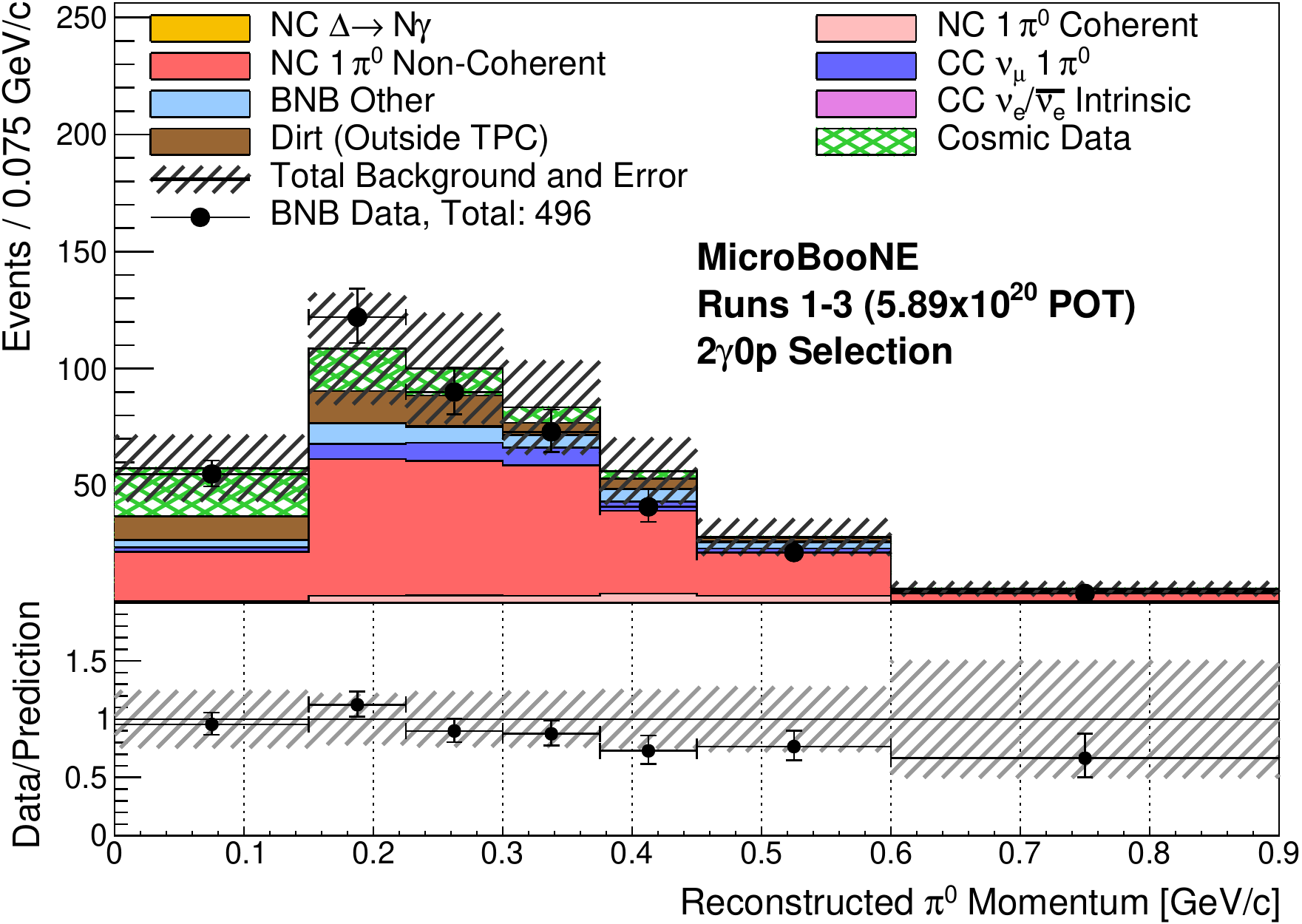}}
    \caption{Data and MC comparisons of the reconstructed $\pi^0$ momentum distributions for the (a) 2$\gamma$1p and (b) 2$\gamma$0p  selected events. }
    \label{fig:ncpi0_finalSel}
\end{figure}

The number of predicted background (both from simulation and cosmic ray data) events and NC $\Delta\rightarrow N\gamma$ signal events after BDT selection are summarized in Table~\ref{tab:events}. A significant background to the search for single-photon events is NC $\pi^0$ events in which one of the decay photons is not reconstructed. 
This happens for a variety of reasons: 
(a) one of the photons from the $\pi^0$ decay may leave the detector active TPC volume before interacting,
(b) the $\pi^0$ decay may be highly asymmetric leading to a secondary photon that is low in energy and not reconstructed, 
(c) both photons may be approximately co-linear and overlapping and thus reconstructed as a single shower, or 
(d) the secondary photon may fall in a region of unresponsive wires, leading to poor reconstruction efficiency. 
Motivated by the background contribution of NC $\pi^0$ events, the $2\gamma1p$ and $2\gamma0p$ event samples serve to constrain the rate of NC $\pi^0$ background. The $2\gamma$ samples follow the same topological, pre-selection and BDT selection scheme as the $1\gamma$ samples (see supplemental materials). The selected $2\gamma1p$ and $2\gamma0p$ events are shown in Fig.~\ref{fig:ncpi0_finalSel} as a function of reconstructed $\pi^0$ momentum, with a true NC $1\pi^0$ event purity of 63.4\% and 59.6\%, respectively. The data-to-MC simulation ratio in the $2\gamma1p$ and $2\gamma0p$ samples is $0.80\pm 0.22$(stat.$\oplus$sys.) and $0.91\pm 0.19$(stat.$\oplus$sys.), respectively, showing an overall deficit but one that is within $1\sigma$.

\begin{table}[t!]
    \centering
    \begin{tabular}{lrr}
    \hline\hline
        Process & $1\gamma1p$ & $1\gamma0p$ \\ \hline
        NC $1\pi^0$ Non-Coherent & 24.0 & 68.1 \\
        NC $1\pi^0$ Coherent & 0.0 & 7.6 \\
        CC $\nu_\mu$ $1\pi^0$ & 0.5 & 14.0\\
        CC $\nu_e$ and $\bar{\nu}_e$ & 0.4 & 11.1\\
        BNB Other & 2.1 &  18.1 \\
        Dirt (outside TPC) & 0.0 & 36.4 \\
        Cosmic Ray Data& 0.0 & 10.0 \\
        \hline
        Total Background (Unconstr.) & 27.0 & 165.4 \\ 
        NC $\Delta\rightarrow N\gamma$ & 4.88 & 6.55 \\ 

        \hline\hline
    \end{tabular}
    \caption{The expected event rates in the $1\gamma1p$ and $1\gamma0p$ samples. ``Dirt (outside TPC)'' represents any neutrino interaction that originates outside the active TPC, but scatters inside. Relative to topological selection stage, the $\nu_e$ CC rejection is 99.8\% and 87.6\% for $1\gamma1p$ and $1\gamma0p$, respectively.}
    \label{tab:events}
\end{table}

The selected data and MC predictions are compared in a fit with a single free parameter corresponding to the normalization ($x_\Delta$) of the nominal rate of NC $\Delta \rightarrow N\gamma$. A single bin is used for each of the $1\gamma1p$ and $1\gamma0p$ event samples, with reconstructed shower energy bin boundaries of 0-600~MeV and 100-700~MeV, respectively. The one-bin $1\gamma1p$ and $1\gamma0p$ event rates are fit simultaneously with the $2\gamma1p$ and $2\gamma0p$ distributions shown in Fig.~\ref{fig:ncpi0_finalSel}. 
The fit makes use of a covariance matrix that encapsulates statistical and systematic uncertainties and bin-to-bin correlations, allowing for both the expected rate and uncertainties of the NC $\pi^0$ backgrounds in the $1\gamma$ samples to be effectively constrained by the high-statistics data observed in the $2\gamma$ samples. 

The normalization $x_\Delta$ can also be reinterpreted in various ways. First, it can be reinterpreted as a scaling of an effective branching fraction $\mathcal{B}_\text{eff}(\Delta\rightarrow N\gamma)$, where the nominal prediction ($x_\Delta=1$) corresponds to an effective branching fraction of 0.6\%. 
This effective branching fraction can be thought of as a metric to account for any uncertain nuclear effects that might modify the $\Delta$ behavior inside the nuclear medium, as we cannot observe the true $\Delta\rightarrow N\gamma$ branching fraction directly. In addition, any BSM effect that can contribute as an NC $\Delta$-like process (with a single photon-like shower in the final state) could lead to an effective modification to the observed branching fraction. Although \textsc{genie} prescribes a normalization uncertainty for $\mathcal{B}_\text{eff}(\Delta\rightarrow N\gamma)$, this uncertainty is not included in the fit. In addition, with the knowledge that \textsc{genie} predicts a cross-section for NC $\Delta\rightarrow N\gamma$ production to be $\sigma^\text{GENIE,Ar}_{NC \Delta\rightarrow N\gamma} = 8.61\times10^{-42}$cm$^{-2}$/nucleon, we can also reinterpret $x_\Delta$ as scaling on this production cross-section. The Feldman-Cousins \cite{Feldman:1997qc} approach is followed to construct the confidence intervals for $x_\Delta$ given the best fit to the observed data, with a metric of $\Delta \chi^2$ defined using the Combined-Neyman-Pearson $\chi^2$ \cite{Ji_2020} as an approximation of the log-likelihood ratio. 

Systematic uncertainties include contributions from flux, cross-section modeling, hadron re-interactions, detector effects, and finite statistics used in the background predictions (both MC and cosmic ray data). The flux uncertainties incorporate hadron production uncertainties, uncertainties on pion and nucleon scattering in the beryllium target and surrounding aluminum magnetic horn, and mis-modeling of the horn current. Following \cite{MicroBooNE:2019nio}, 
these are implemented by reweighting the flux prediction and studying the propagated effects on event distributions. 
The cross-section uncertainties incorporate modeling uncertainties on the \textsc{genie} prediction~\cite{MicroBooNE:2021ccs,andreopoulos2010genie,genie_v3}, evaluated also by reweighting tools. 
The hadron-argon re-interaction uncertainties are associated with the propagation of hadrons through the detector, as modeled in \textsc{geant4} \cite{geant4}. 
The detector modeling and response uncertainties 
are evaluated using a novel data-driven technique. This uses \textit{in-situ} measurements of distortions in the TPC wire readout signals due to various detector effects, such as diffusion, electron drift lifetime, electric field, and electronics response, to parametrize these effects at the TPC wire level, and provides a detector model-agnostic way to study and evaluate their effects on event distributions~\cite{MicroBooNE:2021roa}. Additional systematics varying the charge recombination model, the scintillation light yield, and space charge effects~\cite{yifan_sce,mike_sce} are separately included.  The uncertainty on photo-nuclear absorption of photons on argon was evaluated to be at the sub-percent level, and is therefore omitted. There is also no assigned uncertainty for heavier resonances or coherent single-photon production, which are not simulated in \textsc{genie}. Finally, an inconsistency was identified in the \textsc{genie} v3.0.6 reweighing code used to evaluate a small subset of systematic uncertainties, but was found to have negligible impact on the analysis sensitivity and thus has been ignored.

 The fractional systematic uncertainties on the $1\gamma1p$ and $1\gamma0p$ total background events are summarized in Table~\ref{tab:sys}. The \textsc{genie} cross-section uncertainties dominate. This stems from the uncertainties on NC $\pi^0$ production on argon, which forms the largest background and has not been measured to high precision to date. 
 Both cross-section and flux uncertainties are strongly correlated between the $1\gamma$ and $2\gamma$ event samples. The simultaneous fit to the $1\gamma$ and $2\gamma$ samples is equivalent to a $1\gamma$-only fit where the background and uncertainty are conditionally constrained \cite{alma991015286869705251} by the $2\gamma$ samples.   
 Given the $2\gamma$ samples' statistics, this constraint effectively reduces the total background systematic uncertainty of the $1\gamma1p$ and $1\gamma0p$ samples by 40\% and 50\%, and the total background prediction by 24.1\% and 12.3\%, respectively. 

 \begin{table}[!h]
    \centering
    \begin{tabular}{lrr}
    \hline\hline
         Type of Uncertainty & $1\gamma1p$  & $1\gamma0p$  \\ \hline 
         Flux model  & 7.4\% &  6.6\%\\
         \textsc{genie} cross-section model  & 24.8\% & 16.3\% \\
         \textsc{geant4} re-interactions  & 1.1\% & 1.3\%  \\ 
         Detector effects  & 12.2\% & 6.4\%  \\
         Finite background statistics & 8.3\% & 4.0\% \\ \hline 
         Total Uncertainty (Unconstr.) & 29.8\% & 19.2\% \\ \hline 
         Total Uncertainty (Constr.) & 17.8\% & 9.5\% \\ \hline \hline
    \end{tabular}
    \caption{
    Breakdown of background systematic uncertainties for the $1\gamma1p$ and $1\gamma0p$ samples.
    }
    \label{tab:sys}
\end{table}

The 90\% CL sensitivity is quantified for a Feldman-Cousins-corrected limit in the case of a background-only observation, $x_\Delta=0$, to be less than $x_\Delta=2.5$, corresponding to $\mathcal{B}_\text{eff}(\Delta\rightarrow N\gamma)=1.50\%$  and $\sigma^\text{Ar}_{NC \Delta\rightarrow N\gamma} = 21.5\times10^{-42}$cm$^{-2}$/nucleon . Under a two-hypothesis $\Delta\chi^2$ test, the expected sensitivity of the median experiment assuming the nominal prediction, to reject the LEE hypothesis ($x_\text{MB}=3.18$) in favor of the nominal hypothesis ($x_\Delta=1$) is 1.5$\sigma$; in the case of the median experiment assuming the LEE hypothesis, the sensitivity to reject the nominal hypothesis in favor of the LEE hypothesis is 1.6$\sigma$.

\begin{figure}[b!]
    \subfloat[]{
        \includegraphics[trim={0 0cm 0 0},clip,width=\linewidth]{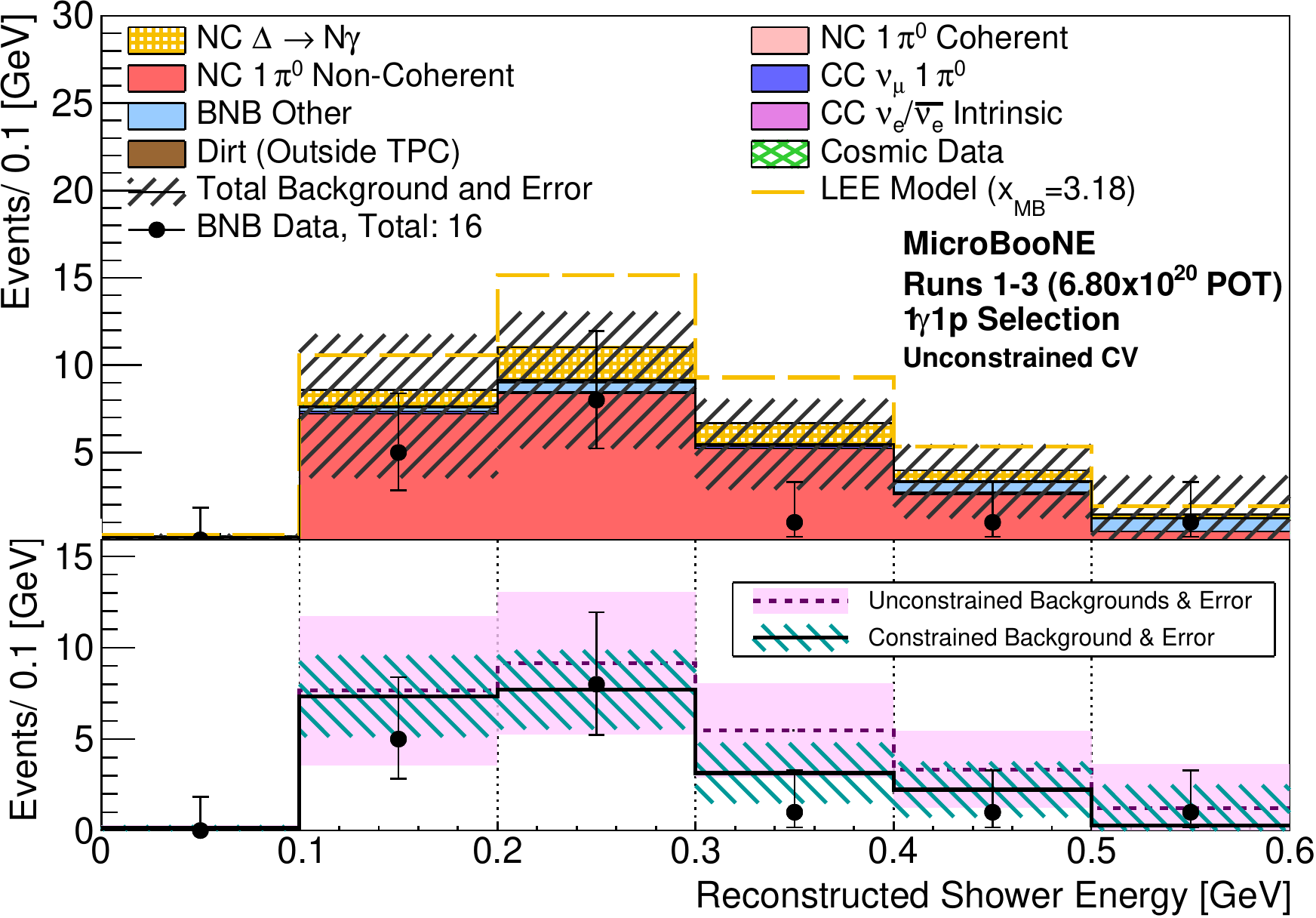}
        }\\
   \subfloat[]{
        \includegraphics[trim={0 0cm 0 0},clip,width=\linewidth]{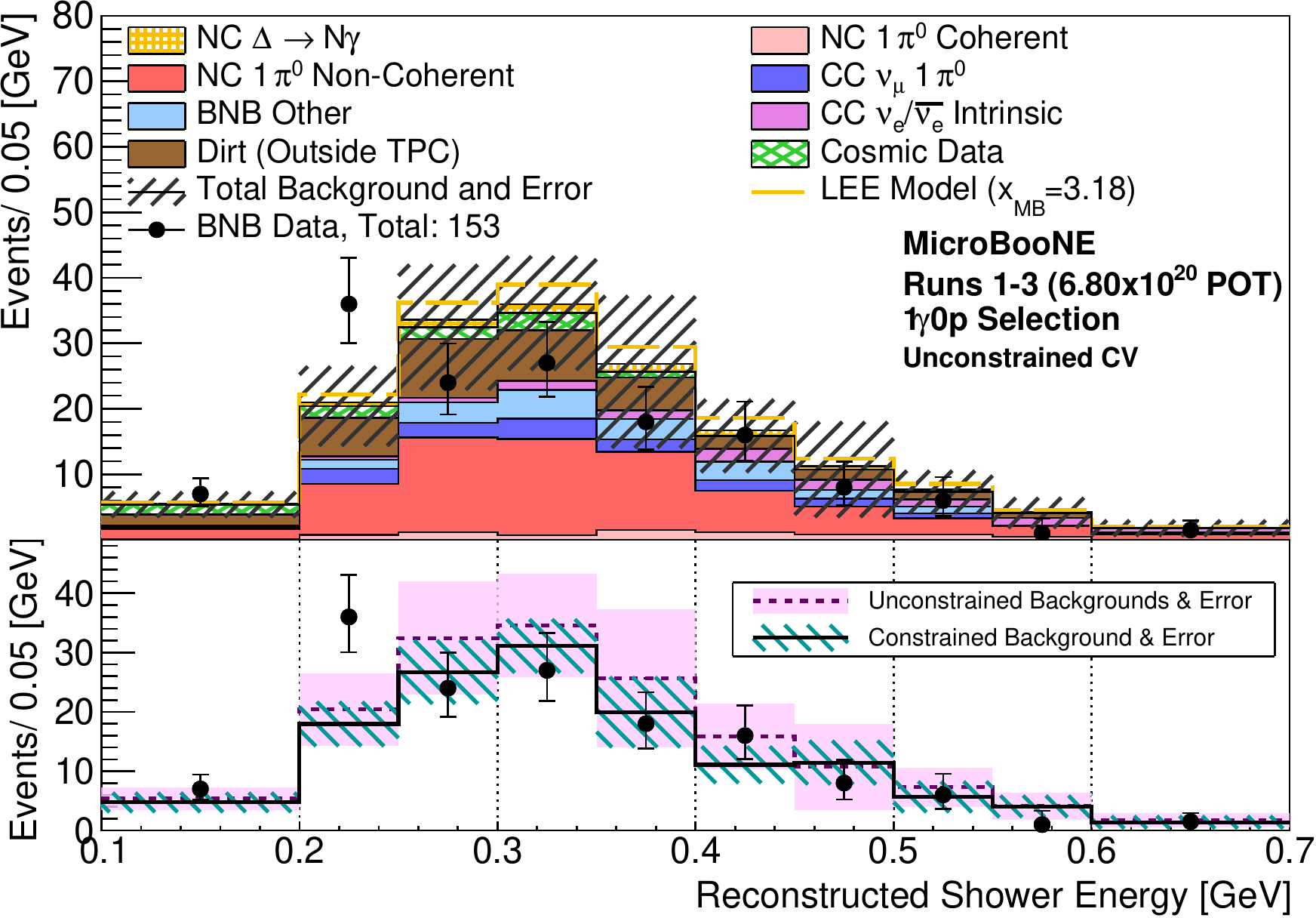}
       }
    \caption{Energy spectra for the (a) $1\gamma1p$ and (b) $1\gamma0p$ selected events.  The upper section in each figure shows the unconstrained background predictions and breakdowns as a function of reconstructed shower energy. The lower section shows the total background prediction with systematic uncertainty both before and after the $2\gamma$ constraint. The local significance of the data fluctuation in the 200-250~MeV bin of (b) corresponds to $1.6\sigma$ ($\chi^2/dof = 3.66/1$) before the $2\gamma$ constraint, and $2.7\sigma$ ($\chi^2/dof = 8.54/1$) after. From MC studies, the probability of any one bin across all 16 $1\gamma$ bins giving rise to a constrained $\chi^2 \ge 8.54$ is 4.74\%.
 }
    \label{fig:1g_finalSel}
\end{figure}

\begin{figure}[tb]
    \centering
    \vspace{-0.5cm}
    \subfloat[]{
        \includegraphics[width=\linewidth]{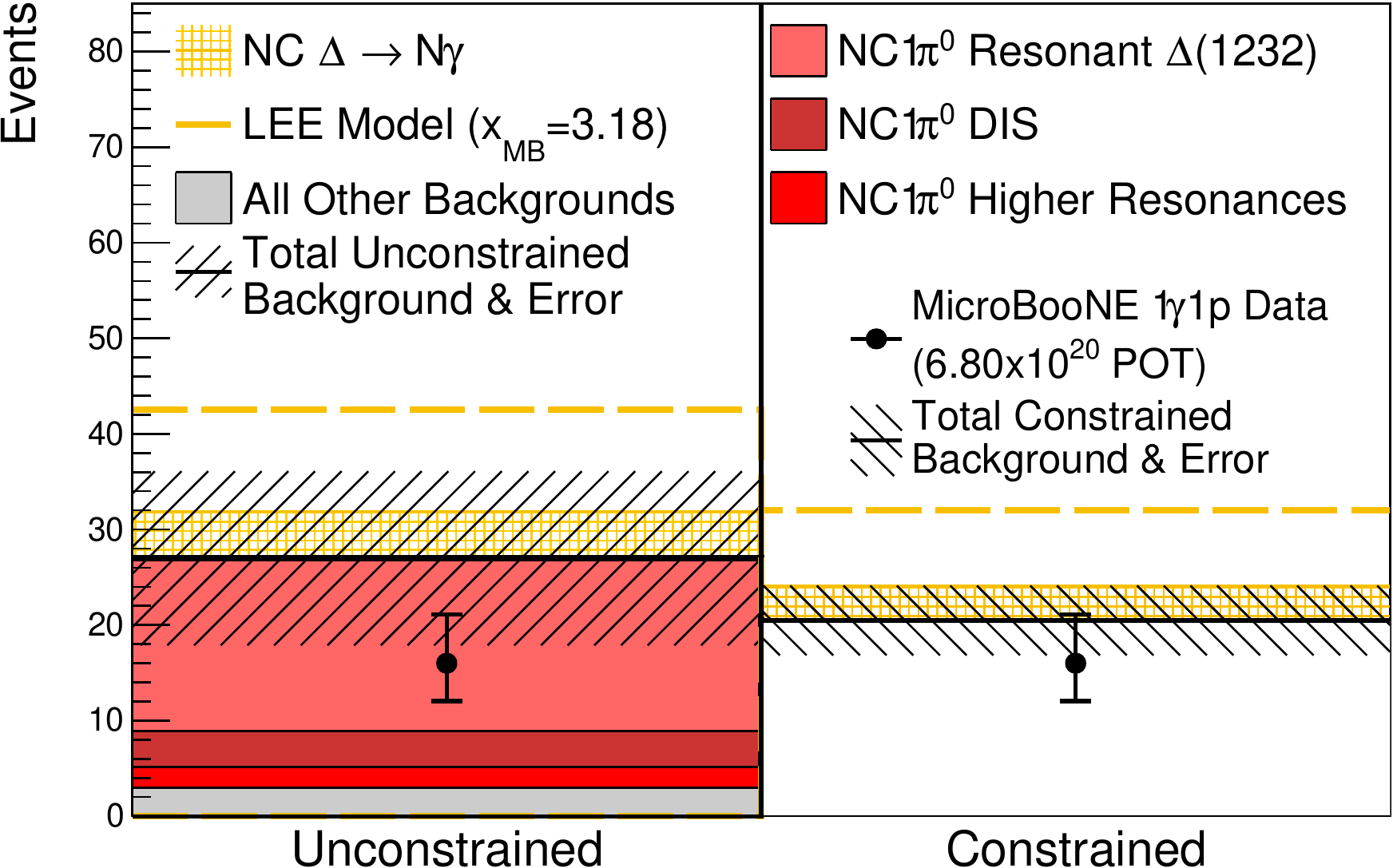}
    }\\
    \subfloat[]{
        \includegraphics[width=\linewidth]{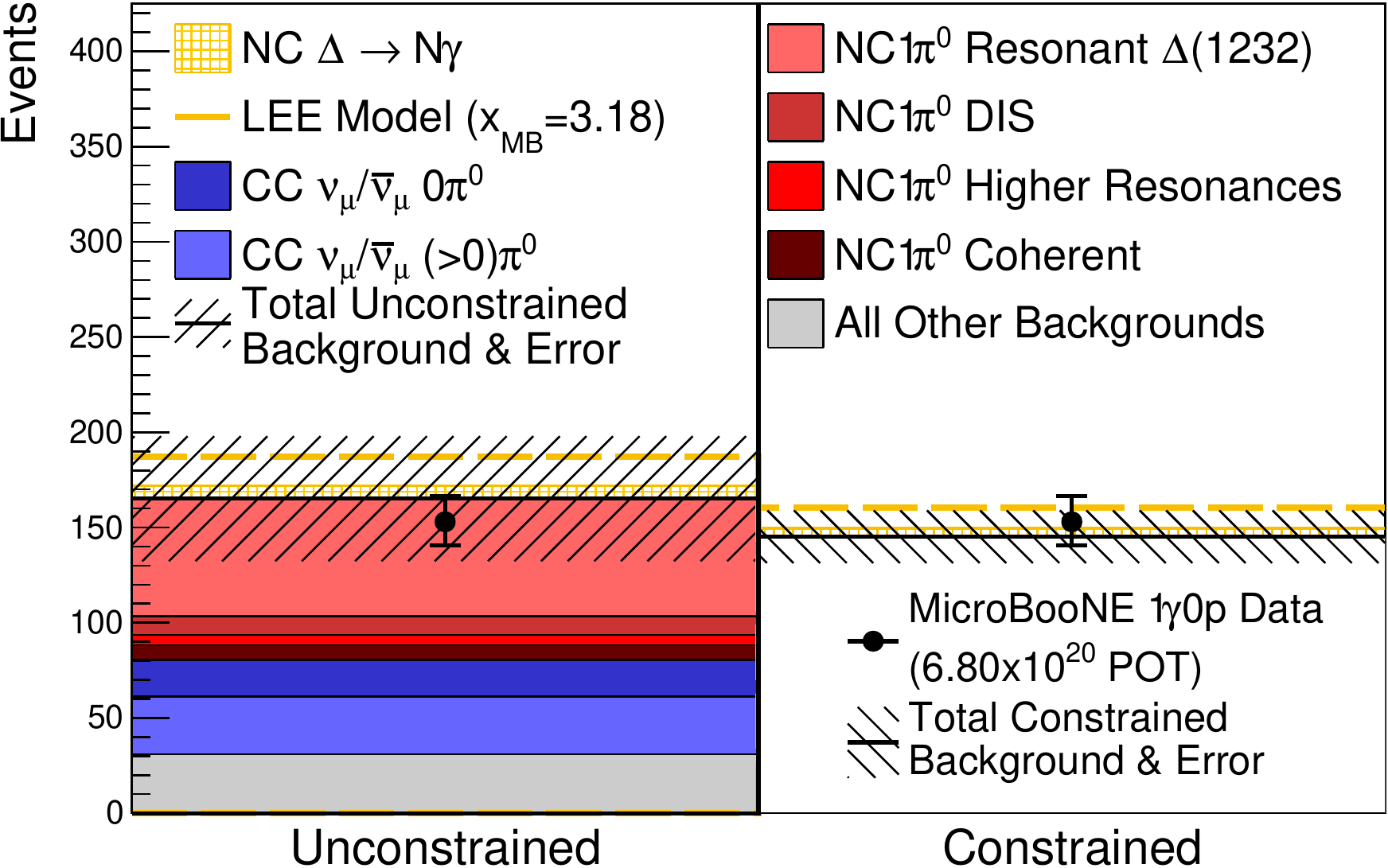}
        }
    \caption{The observed event rates for the (a) $1\gamma1p$ and (b) $1\gamma0p$ event samples, and comparisons to unconstrained (left) and constrained (right) background and LEE model predictions. The event rates are the sum of all events with reconstructed shower energy between 0-600~MeV and 100-700~MeV for (a) and (b), respectively. The one-bin background only conditionally constrained $\chi^2$ is 0.63 and 0.18 for $1\gamma1p$ and $1\gamma0p$ respectively. 
    }
    \label{fig:results}
\end{figure}

The reconstruction, selection, and fitting methods employed in this search were developed adhering to a signal-blind analysis strategy, whereby the data was kept blind until the analysis was fully developed, with the exception of a small subset of the data consisting of $0.51\times10^{20}$~POT, used for analysis validation.  After $1\gamma1p$ and $1\gamma0p$ event samples were unblinded, 16 data events with an expected constrained background of $20.5  \pm 3.6 \text{(sys.)} $ events were observed in the $1\gamma1p$ event sample, and 153 data events with an expected constrained background of $145.1  \pm 13.8 \text{(sys.)} $ events were observed in the $1\gamma0p$ event sample. The reconstructed shower energy distributions of selected $1\gamma1p$ and $1\gamma0p$ events are shown in Fig.~\ref{fig:1g_finalSel}. Overall, a systematic deficit of data relative to the unconstrained MC prediction is observed, which is within systematic and statistical uncertainties, and consistent with a similar deficit in the $2\gamma$ event samples. The expected signal and background predictions are summarized in Table~\ref{tab:results} and Fig.~\ref{fig:results}, and compared to the observed data, both before and after applying the $2\gamma$ conditional constraint. 
The $2\gamma$ constraint reduces the total background prediction, consistently with the data to MC simulation ratio observed in the $2\gamma$ event samples. 

\begin{table}[t!]
    \centering
    \begin{tabular}{lcccc}
    \hline\hline
    & $1\gamma1p$ && $1\gamma0p$ \\ \hline\hline
    Unconstr. bkgd. & 27.0 $ \pm $ 8.1    && 165.4 $ \pm $ 31.7 \\
    Constr. bkgd. & 20.5 $ \pm $ 3.6   && 145.1 $ \pm $ 13.8\\
    \hline
    NC $\Delta\rightarrow N\gamma$ & 4.88 && 6.55 \\
    LEE ($x_\text{MB}=3.18$) & 15.5 && 20.1 \\
    \hline 
    Data & 16 &&153 \\
    \hline\hline
    \end{tabular}
    \caption{Number of predicted background, predicted signal, and observed data events for the $1\gamma1p$ and $1\gamma0p$ samples, with background systematic uncertainties.}
    \label{tab:results}
\end{table}

\begin{figure}[!h]
    \centering 
        \includegraphics[width=\linewidth, trim=0 0 0 0]{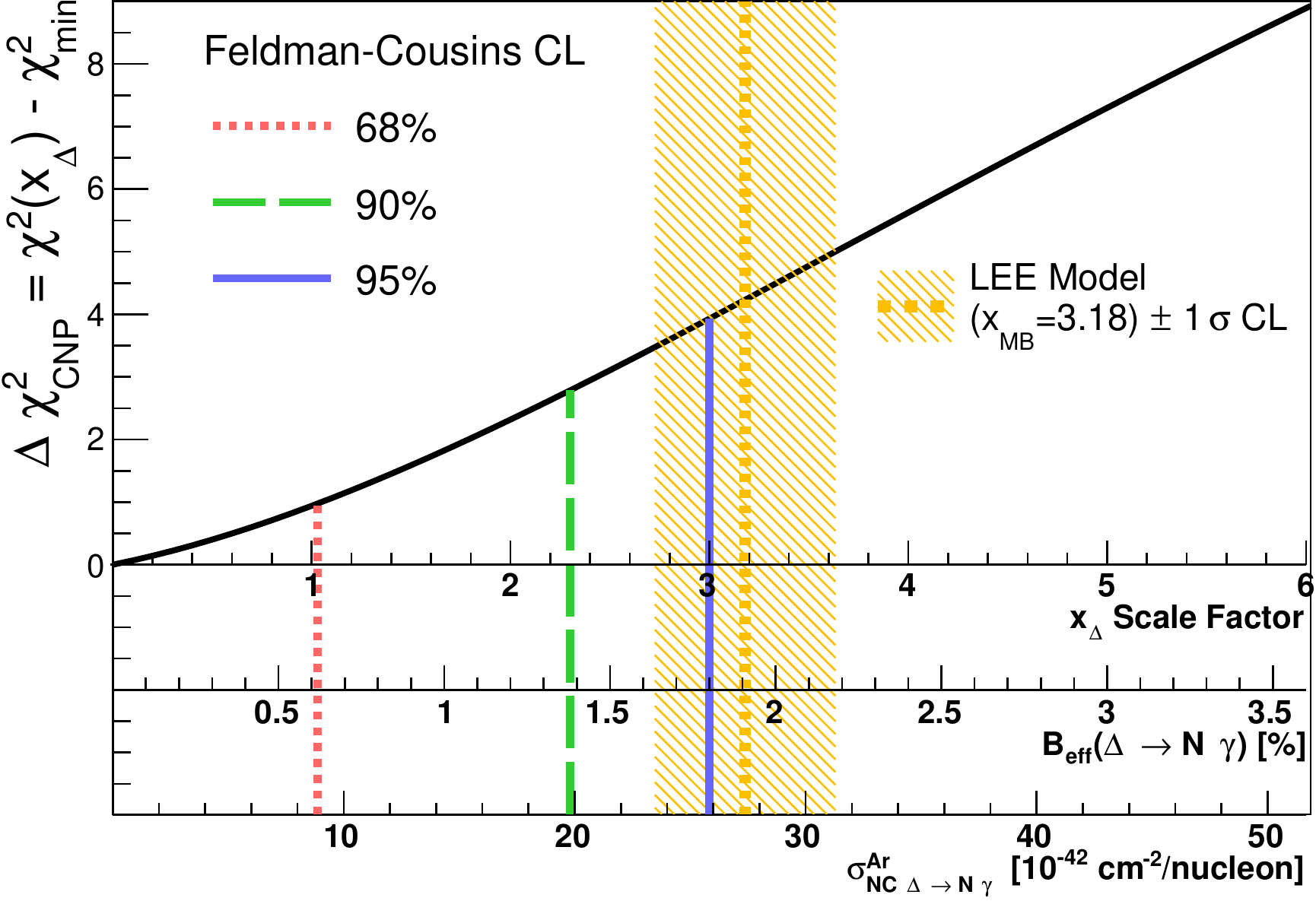}
    \caption{ 
    The resulting $\Delta \chi^2$ curve after fitting to $x_\Delta$ using the Feldman-Cousins procedure, showing extracted confidence intervals. The best fit is found to be at $x_\Delta = 0$ with a $\chi^2_{bf} =5.53$. Shown also is the reinterpretation of this scaling factor as both an effective branching fraction, $\mathcal{B}_\text{eff}(\Delta\rightarrow N\gamma)$, and a cross-section, $\sigma^\text{Ar}_{NC \Delta\rightarrow N\gamma}$. The default \textsc{genie} value corresponds to $x_\Delta=1$. The error on the LEE model is estimated from the MiniBooNE result \cite{Aguilar_Arevalo_2021} with statistical and systematic uncertainty. It should be noted that this uncertainty does not account for systematic correlations between MiniBooNE and MicroBooNE.}
    \label{fig:delta_chi}
\end{figure}

The best-fit value for $x_{\Delta}$ obtained from the fit is 0, with a $\chi^2_{bf}$ of 5.53 for 15 degrees of freedom ($dof$). This measurement is in agreement with the nominal NC $\Delta \rightarrow N \gamma$ rate (corresponding to $x_\Delta=1$) within $1\sigma$ (67.8\%~CL) with a $\chi^2$ of 6.47 for 16 $dof$. The Feldman-Cousins calculated confidence limit leads to a one-sided bound on the normalization of NC $\Delta\rightarrow N\gamma$ events of $x_{\Delta} < 2.3$, corresponding to $\mathcal{B}_\text{eff}(\Delta\rightarrow N\gamma) < 1.38\%$ and $\sigma^\text{Ar}_{NC \Delta\rightarrow N\gamma} < 19.8\times10^{-42}$cm$^{-2}$/nucleon, at 90\%~CL. This is summarized in Fig.~\ref{fig:delta_chi}.

This result represents the most stringent limit on neutrino-induced NC $\Delta \rightarrow N \gamma$ on any nuclear target~\cite{Kullenberg_2012,Abe_2019}, and a significant improvement over previous searches, in particular in the neutrino energy range below 1~GeV. Under a two-hypothesis test, the data rules out the interpretation of the MiniBooNE anomalous excess \cite{mb} as a factor of 3.18 enhancement to the rate of $\Delta\rightarrow N\gamma$, in favor of the nominal prediction at $94.8$\%~CL (1.9$\sigma$). 
 While this is a model-dependent test of the MiniBooNE LEE, and does not apply universally to all other photon-like interpretations, it provides an important constraint on this process and a first direct test of the MiniBooNE LEE, and opens the door to further searches that focus on a broader range of models. Those include coherent single-photon production~\cite{Wang_2014}, anomalous contributions of which could give rise to additional events and would be expected to leave an imprint in the $1\gamma0p$ selection, as well as more exotic beyond-SM processes that manifest as single-photon events, such as co-linear $e^+e^-$ pairs from $Z'$~\cite{Bertuzzo:2018itn,Ballett:2018ynz} or scalar~\cite{Abdallah:2020vgg} decays, among others. Follow-up MicroBooNE analyses will explicitly target and quantify sensitivity to these alternative hypotheses, as well as model-independent single-photon searches.

\begin{acknowledgments}
This document was prepared by the MicroBooNE collaboration using the
resources of the Fermi National Accelerator Laboratory (Fermilab), a
U.S. Department of Energy, Office of Science, HEP User Facility.
Fermilab is managed by Fermi Research Alliance, LLC (FRA), acting
under Contract No. DE-AC02-07CH11359.  MicroBooNE is supported by the
following: the U.S. Department of Energy, Office of Science, Offices
of High Energy Physics and Nuclear Physics; the U.S. National Science
Foundation; the Swiss National Science Foundation; the Science and
Technology Facilities Council (STFC), part of the United Kingdom Research 
and Innovation; the Royal Society (United Kingdom); and The European 
Union’s Horizon 2020 Marie Skłodowska-Curie Actions. Additional support 
for the laser calibration system and cosmic ray tagger was provided by 
the Albert Einstein Center for Fundamental Physics, Bern, Switzerland.

\end{acknowledgments}

\bibliography{apssamp}

\begin{thebibliography}{32}%
\makeatletter
\providecommand \@ifxundefined [1]{%
 \@ifx{#1\undefined}
}%
\providecommand \@ifnum [1]{%
 \ifnum #1\expandafter \@firstoftwo
 \else \expandafter \@secondoftwo
 \fi
}%
\providecommand \@ifx [1]{%
 \ifx #1\expandafter \@firstoftwo
 \else \expandafter \@secondoftwo
 \fi
}%
\providecommand \natexlab [1]{#1}%
\providecommand \enquote  [1]{``#1''}%
\providecommand \bibnamefont  [1]{#1}%
\providecommand \bibfnamefont [1]{#1}%
\providecommand \citenamefont [1]{#1}%
\providecommand \href@noop [0]{\@secondoftwo}%
\providecommand \href [0]{\begingroup \@sanitize@url \@href}%
\providecommand \@href[1]{\@@startlink{#1}\@@href}%
\providecommand \@@href[1]{\endgroup#1\@@endlink}%
\providecommand \@sanitize@url [0]{\catcode `\\12\catcode `\$12\catcode
  `\&12\catcode `\#12\catcode `\^12\catcode `\_12\catcode `\%12\relax}%
\providecommand \@@startlink[1]{}%
\providecommand \@@endlink[0]{}%
\providecommand \url  [0]{\begingroup\@sanitize@url \@url }%
\providecommand \@url [1]{\endgroup\@href {#1}{\urlprefix }}%
\providecommand \urlprefix  [0]{URL }%
\providecommand \Eprint [0]{\href }%
\providecommand \doibase [0]{http://dx.doi.org/}%
\providecommand \selectlanguage [0]{\@gobble}%
\providecommand \bibinfo  [0]{\@secondoftwo}%
\providecommand \bibfield  [0]{\@secondoftwo}%
\providecommand \translation [1]{[#1]}%
\providecommand \BibitemOpen [0]{}%
\providecommand \bibitemStop [0]{}%
\providecommand \bibitemNoStop [0]{.\EOS\space}%
\providecommand \EOS [0]{\spacefactor3000\relax}%
\providecommand \BibitemShut  [1]{\csname bibitem#1\endcsname}%
\let\auto@bib@innerbib\@empty
\bibitem [{\citenamefont {Aguilar-Arevalo}\ \emph
  {et~al.}(2009{\natexlab{a}})\citenamefont {Aguilar-Arevalo} \emph
  {et~al.}}]{MiniBooNE:2008yuf}%
  \BibitemOpen
  \bibfield  {author} {\bibinfo {author} {\bibfnamefont {A.~A.}\ \bibnamefont
  {Aguilar-Arevalo}} \emph {et~al.} (\bibinfo {collaboration} {MiniBooNE}),\
  }\href {\doibase 10.1103/PhysRevLett.102.101802} {\bibfield  {journal}
  {\bibinfo  {journal} {Phys. Rev. Lett.}\ }\textbf {\bibinfo {volume} {102}},\
  \bibinfo {pages} {101802} (\bibinfo {year} {2009}{\natexlab{a}})},\ \Eprint
  {http://arxiv.org/abs/0812.2243} {arXiv:0812.2243 [hep-ex]} \BibitemShut
  {NoStop}%
\bibitem [{\citenamefont {Aguilar-Arevalo}\ \emph {et~al.}(2018)\citenamefont
  {Aguilar-Arevalo} \emph {et~al.}}]{MiniBooNE:2018esg}%
  \BibitemOpen
  \bibfield  {author} {\bibinfo {author} {\bibfnamefont {A.~A.}\ \bibnamefont
  {Aguilar-Arevalo}} \emph {et~al.} (\bibinfo {collaboration} {MiniBooNE
  Collaboration}),\ }\href {\doibase 10.1103/PhysRevLett.121.221801} {\bibfield
   {journal} {\bibinfo  {journal} {Phys. Rev. Lett.}\ }\textbf {\bibinfo
  {volume} {121}},\ \bibinfo {pages} {221801} (\bibinfo {year}
  {2018})}\BibitemShut {NoStop}%
\bibitem [{\citenamefont {Aguilar-Arevalo}\ \emph {et~al.}(2021)\citenamefont
  {Aguilar-Arevalo} \emph {et~al.}}]{Aguilar_Arevalo_2021}%
  \BibitemOpen
  \bibfield  {author} {\bibinfo {author} {\bibfnamefont {A.~A.}\ \bibnamefont
  {Aguilar-Arevalo}} \emph {et~al.} (\bibinfo {collaboration} {MiniBooNE
  Collaboration}),\ }\href {\doibase 10.1103/PhysRevD.103.052002} {\bibfield
  {journal} {\bibinfo  {journal} {Phys. Rev. D}\ }\textbf {\bibinfo {volume}
  {103}},\ \bibinfo {pages} {052002} (\bibinfo {year} {2021})}\BibitemShut
  {NoStop}%
\bibitem [{\citenamefont {Athanassopoulos}\ \emph {et~al.}(1996)\citenamefont
  {Athanassopoulos} \emph {et~al.}}]{LSND:1996vlr}%
  \BibitemOpen
  \bibfield  {author} {\bibinfo {author} {\bibfnamefont {C.}~\bibnamefont
  {Athanassopoulos}} \emph {et~al.} (\bibinfo {collaboration} {LSND}),\ }\href
  {\doibase 10.1103/PhysRevC.54.2685} {\bibfield  {journal} {\bibinfo
  {journal} {Phys. Rev. C}\ }\textbf {\bibinfo {volume} {54}},\ \bibinfo
  {pages} {2685} (\bibinfo {year} {1996})},\ \Eprint
  {http://arxiv.org/abs/nucl-ex/9605001} {arXiv:nucl-ex/9605001} \BibitemShut
  {NoStop}%
\bibitem [{\citenamefont {Wang}\ \emph {et~al.}(2014)\citenamefont {Wang},
  \citenamefont {Alvarez-Ruso},\ and\ \citenamefont {Nieves}}]{Wang_2014}%
  \BibitemOpen
  \bibfield  {author} {\bibinfo {author} {\bibfnamefont {E.}~\bibnamefont
  {Wang}}, \bibinfo {author} {\bibfnamefont {L.}~\bibnamefont {Alvarez-Ruso}},
  \ and\ \bibinfo {author} {\bibfnamefont {J.}~\bibnamefont {Nieves}},\ }\href
  {\doibase 10.1103/PhysRevC.89.015503} {\bibfield  {journal} {\bibinfo
  {journal} {Phys. Rev. C}\ }\textbf {\bibinfo {volume} {89}},\ \bibinfo
  {pages} {015503} (\bibinfo {year} {2014})}\BibitemShut {NoStop}%
\bibitem [{\citenamefont {Blanpied}\ \emph {et~al.}(1997)\citenamefont
  {Blanpied} \emph {et~al.}}]{Blanpied:1997zz}%
  \BibitemOpen
  \bibfield  {author} {\bibinfo {author} {\bibfnamefont {G.}~\bibnamefont
  {Blanpied}} \emph {et~al.},\ }\href {\doibase 10.1103/PhysRevLett.79.4337}
  {\bibfield  {journal} {\bibinfo  {journal} {Phys. Rev. Lett.}\ }\textbf
  {\bibinfo {volume} {79}},\ \bibinfo {pages} {4337} (\bibinfo {year}
  {1997})}\BibitemShut {NoStop}%
\bibitem [{\citenamefont {Sparveris}\ \emph {et~al.}(2008)\citenamefont
  {Sparveris} \emph {et~al.}}]{Sparveris:2008jx}%
  \BibitemOpen
  \bibfield  {author} {\bibinfo {author} {\bibfnamefont {N.~F.}\ \bibnamefont
  {Sparveris}} \emph {et~al.},\ }\href {\doibase 10.1103/PhysRevC.78.018201}
  {\bibfield  {journal} {\bibinfo  {journal} {Phys. Rev. C}\ }\textbf {\bibinfo
  {volume} {78}},\ \bibinfo {pages} {018201} (\bibinfo {year} {2008})},\
  \Eprint {http://arxiv.org/abs/0804.1169} {arXiv:0804.1169 [nucl-ex]}
  \BibitemShut {NoStop}%
\bibitem [{\citenamefont {Abe}\ \emph {et~al.}(2019)\citenamefont {Abe} \emph
  {et~al.}}]{Abe_2019}%
  \BibitemOpen
  \bibfield  {author} {\bibinfo {author} {\bibfnamefont {K.}~\bibnamefont
  {Abe}} \emph {et~al.} (\bibinfo {collaboration} {T2K}),\ }\href {\doibase
  10.1088/1361-6471/ab227d} {\bibfield  {journal} {\bibinfo  {journal} {J.
  Phys. G: Nucl. Part. Phys.}\ }\textbf {\bibinfo {volume} {46}},\ \bibinfo
  {pages} {08LT01} (\bibinfo {year} {2019})}\BibitemShut {NoStop}%
\bibitem [{\citenamefont {Kullenberg}\ \emph {et~al.}(2012)\citenamefont
  {Kullenberg} \emph {et~al.}}]{Kullenberg_2012}%
  \BibitemOpen
  \bibfield  {author} {\bibinfo {author} {\bibfnamefont {C.}~\bibnamefont
  {Kullenberg}} \emph {et~al.} (\bibinfo {collaboration} {NOMAD}),\ }\href
  {\doibase 10.1016/j.physletb.2011.11.049} {\bibfield  {journal} {\bibinfo
  {journal} {Phys. Lett. B}\ }\textbf {\bibinfo {volume} {706}},\ \bibinfo
  {pages} {268–275} (\bibinfo {year} {2012})}\BibitemShut {NoStop}%
\bibitem [{\citenamefont {Acciarri}\ \emph {et~al.}(2017)\citenamefont
  {Acciarri} \emph {et~al.}}]{Acciarri_2017}%
  \BibitemOpen
  \bibfield  {author} {\bibinfo {author} {\bibfnamefont {R.}~\bibnamefont
  {Acciarri}} \emph {et~al.} (\bibinfo {collaboration} {MicroBooNE}),\
  }\href@noop {} {\bibfield  {journal} {\bibinfo  {journal} {JINST}\ }\textbf
  {\bibinfo {volume} {12}},\ \bibinfo {pages} {P02017} (\bibinfo {year}
  {2017})}\BibitemShut {NoStop}%
\bibitem [{\citenamefont {Aguilar-Arevalo}\ \emph
  {et~al.}(2009{\natexlab{b}})\citenamefont {Aguilar-Arevalo} \emph
  {et~al.}}]{PhysRevD.79.072002}%
  \BibitemOpen
  \bibfield  {author} {\bibinfo {author} {\bibfnamefont {A.~A.}\ \bibnamefont
  {Aguilar-Arevalo}} \emph {et~al.} (\bibinfo {collaboration} {MiniBooNE}),\
  }\href {\doibase 10.1103/PhysRevD.79.072002} {\bibfield  {journal} {\bibinfo
  {journal} {Phys. Rev. D}\ }\textbf {\bibinfo {volume} {79}},\ \bibinfo
  {pages} {072002} (\bibinfo {year} {2009}{\natexlab{b}})}\BibitemShut
  {NoStop}%
\bibitem [{\citenamefont {Casper}(2002)}]{nuance-miniboone}%
  \BibitemOpen
  \bibfield  {author} {\bibinfo {author} {\bibfnamefont {D.}~\bibnamefont
  {Casper}},\ }\href {\doibase https://doi.org/10.1016/S0920-5632(02)01756-5}
  {\bibfield  {journal} {\bibinfo  {journal} {Nucl. Phys. B - Proc. Suppl.}\
  }\textbf {\bibinfo {volume} {112}},\ \bibinfo {pages} {161} (\bibinfo {year}
  {2002})}\BibitemShut {NoStop}%
\bibitem [{\citenamefont {Abratenko}\ \emph
  {et~al.}(2021{\natexlab{a}})\citenamefont {Abratenko} \emph
  {et~al.}}]{MicroBooNE:2021ccs}%
  \BibitemOpen
  \bibfield  {author} {\bibinfo {author} {\bibfnamefont {P.}~\bibnamefont
  {Abratenko}} \emph {et~al.} (\bibinfo {collaboration} {MicroBooNE}),\
  }\href@noop {} {\  (\bibinfo {year} {2021}{\natexlab{a}})},\ \Eprint
  {http://arxiv.org/abs/2110.14028} {arXiv:2110.14028 [hep-ex]} \BibitemShut
  {NoStop}%
\bibitem [{\citenamefont {Andreopoulos}\ \emph
  {et~al.}(2010{\natexlab{a}})\citenamefont {Andreopoulos} \emph
  {et~al.}}]{Andreopoulos:2009rq}%
  \BibitemOpen
  \bibfield  {author} {\bibinfo {author} {\bibfnamefont {C.}~\bibnamefont
  {Andreopoulos}} \emph {et~al.},\ }\href {\doibase 10.1016/j.nima.2009.12.009}
  {\bibfield  {journal} {\bibinfo  {journal} {Nucl. Instrum. Meth. A}\ }\textbf
  {\bibinfo {volume} {614}},\ \bibinfo {pages} {87} (\bibinfo {year}
  {2010}{\natexlab{a}})},\ \Eprint {http://arxiv.org/abs/0905.2517}
  {arXiv:0905.2517 [hep-ph]} \BibitemShut {NoStop}%
\bibitem [{\citenamefont {Tena-Vidal}\ \emph {et~al.}(2021)\citenamefont
  {Tena-Vidal} \emph {et~al.}}]{genie_v3}%
  \BibitemOpen
  \bibfield  {author} {\bibinfo {author} {\bibfnamefont {J.}~\bibnamefont
  {Tena-Vidal}} \emph {et~al.} (\bibinfo {collaboration} {GENIE
  Collaboration}),\ }\href@noop {} {\bibfield  {journal} {\bibinfo  {journal}
  {arXiv:2104.09179}\ } (\bibinfo {year} {2021})}\BibitemShut {NoStop}%
\bibitem [{\citenamefont {Alvarez-Ruso}\ \emph {et~al.}(2018)\citenamefont
  {Alvarez-Ruso}, \citenamefont {Nieves}, \citenamefont {Sala},\ and\
  \citenamefont {Wang}}]{coherent}%
  \BibitemOpen
  \bibfield  {author} {\bibinfo {author} {\bibfnamefont {L.}~\bibnamefont
  {Alvarez-Ruso}}, \bibinfo {author} {\bibfnamefont {J.}~\bibnamefont
  {Nieves}}, \bibinfo {author} {\bibfnamefont {E.}~\bibnamefont {Sala}}, \ and\
  \bibinfo {author} {\bibfnamefont {E.}~\bibnamefont {Wang}},\ }\href {\doibase
  10.1088/1742-6596/1056/1/012001} {\bibfield  {journal} {\bibinfo  {journal}
  {Journal of Physics: Conference Series}\ }\textbf {\bibinfo {volume}
  {1056}},\ \bibinfo {pages} {012001} (\bibinfo {year} {2018})}\BibitemShut
  {NoStop}%
\bibitem [{\citenamefont {Acciarri}\ \emph {et~al.}(2018)\citenamefont
  {Acciarri} \emph {et~al.}}]{ub_pandora}%
  \BibitemOpen
  \bibfield  {author} {\bibinfo {author} {\bibfnamefont {R.}~\bibnamefont
  {Acciarri}} \emph {et~al.} (\bibinfo {collaboration} {MicroBooNE}),\ }\href
  {\doibase 10.1140/epjc/s10052-017-5481-6} {\bibfield  {journal} {\bibinfo
  {journal} {Eur. Phys. J. C}\ }\textbf {\bibinfo {volume} {78}},\ \bibinfo
  {pages} {82} (\bibinfo {year} {2018})}\BibitemShut {NoStop}%
\bibitem [{\citenamefont {Chen}\ and\ \citenamefont
  {Guestrin}(2016)}]{Chen:2016}%
  \BibitemOpen
  \bibfield  {author} {\bibinfo {author} {\bibfnamefont {T.}~\bibnamefont
  {Chen}}\ and\ \bibinfo {author} {\bibfnamefont {C.}~\bibnamefont {Guestrin}}\
  }(\bibinfo  {publisher} {Association for Computing Machinery},\ \bibinfo
  {address} {New York, NY, USA},\ \bibinfo {year} {2016})\ pp.\ \bibinfo
  {pages} {785--794}\BibitemShut {NoStop}%
\bibitem [{\citenamefont {Ester}\ \emph {et~al.}(1996)\citenamefont {Ester}
  \emph {et~al.}}]{dbscan}%
  \BibitemOpen
  \bibfield  {author} {\bibinfo {author} {\bibfnamefont {M.}~\bibnamefont
  {Ester}} \emph {et~al.},\ }\href@noop {} {\bibfield  {journal} {\bibinfo
  {journal} {KDD'96: Proceedings of the Second International Conference on
  Knowledge Discovery and Data Mining}\ ,\ \bibinfo {pages} {226}} (\bibinfo
  {year} {1996})}\BibitemShut {NoStop}%
\bibitem [{\citenamefont {Feldman}\ and\ \citenamefont
  {Cousins}(1998)}]{Feldman:1997qc}%
  \BibitemOpen
  \bibfield  {author} {\bibinfo {author} {\bibfnamefont {G.~J.}\ \bibnamefont
  {Feldman}}\ and\ \bibinfo {author} {\bibfnamefont {R.~D.}\ \bibnamefont
  {Cousins}},\ }\href {\doibase 10.1103/PhysRevD.57.3873} {\bibfield  {journal}
  {\bibinfo  {journal} {Phys. Rev. D}\ }\textbf {\bibinfo {volume} {57}},\
  \bibinfo {pages} {3873} (\bibinfo {year} {1998})}\BibitemShut {NoStop}%
\bibitem [{\citenamefont {Ji}\ \emph {et~al.}(2020)\citenamefont {Ji} \emph
  {et~al.}}]{Ji_2020}%
  \BibitemOpen
  \bibfield  {author} {\bibinfo {author} {\bibfnamefont {X.}~\bibnamefont {Ji}}
  \emph {et~al.},\ }\href {\doibase 10.1016/j.nima.2020.163677} {\bibfield
  {journal} {\bibinfo  {journal} {Nucl. Instr. and Meth. A}\ }\textbf {\bibinfo
  {volume} {961}},\ \bibinfo {pages} {163677} (\bibinfo {year}
  {2020})}\BibitemShut {NoStop}%
\bibitem [{\citenamefont {Abratenko}\ \emph {et~al.}(2019)\citenamefont
  {Abratenko} \emph {et~al.}}]{MicroBooNE:2019nio}%
  \BibitemOpen
  \bibfield  {author} {\bibinfo {author} {\bibfnamefont {P.}~\bibnamefont
  {Abratenko}} \emph {et~al.} (\bibinfo {collaboration} {MicroBooNE}),\ }\href
  {\doibase 10.1103/PhysRevLett.123.131801} {\bibfield  {journal} {\bibinfo
  {journal} {Phys. Rev. Lett.}\ }\textbf {\bibinfo {volume} {123}},\ \bibinfo
  {pages} {131801} (\bibinfo {year} {2019})},\ \Eprint
  {http://arxiv.org/abs/1905.09694} {arXiv:1905.09694 [hep-ex]} \BibitemShut
  {NoStop}%
\bibitem [{\citenamefont {Andreopoulos}\ \emph
  {et~al.}(2010{\natexlab{b}})\citenamefont {Andreopoulos} \emph
  {et~al.}}]{andreopoulos2010genie}%
  \BibitemOpen
  \bibfield  {author} {\bibinfo {author} {\bibfnamefont {C.}~\bibnamefont
  {Andreopoulos}} \emph {et~al.},\ }\href {\doibase 10.1016/j.nima.2009.12.009}
  {\bibfield  {journal} {\bibinfo  {journal} {Nucl. Instr. and Meth. A}\
  }\textbf {\bibinfo {volume} {614}},\ \bibinfo {pages} {87} (\bibinfo {year}
  {2010}{\natexlab{b}})}\BibitemShut {NoStop}%
\bibitem [{\citenamefont {Agostinelli}\ \emph {et~al.}(2003)\citenamefont
  {Agostinelli} \emph {et~al.}}]{geant4}%
  \BibitemOpen
  \bibfield  {author} {\bibinfo {author} {\bibfnamefont {S.}~\bibnamefont
  {Agostinelli}} \emph {et~al.},\ }\href@noop {} {\bibfield  {journal}
  {\bibinfo  {journal} {Nucl. Instr. and Meth. A}\ }\textbf {\bibinfo {volume}
  {506}},\ \bibinfo {pages} {250} (\bibinfo {year} {2003})}\BibitemShut
  {NoStop}%
\bibitem [{\citenamefont {Abratenko}\ \emph
  {et~al.}(2021{\natexlab{b}})\citenamefont {Abratenko} \emph
  {et~al.}}]{MicroBooNE:2021roa}%
  \BibitemOpen
  \bibfield  {author} {\bibinfo {author} {\bibfnamefont {P.}~\bibnamefont
  {Abratenko}} \emph {et~al.} (\bibinfo {collaboration} {MicroBooNE}),\
  }\href@noop {} {\  (\bibinfo {year} {2021}{\natexlab{b}})},\ \Eprint
  {http://arxiv.org/abs/2111.03556} {arXiv:2111.03556 [hep-ex]} \BibitemShut
  {NoStop}%
\bibitem [{\citenamefont {Abratenko}\ \emph
  {et~al.}(2020{\natexlab{a}})\citenamefont {Abratenko} \emph
  {et~al.}}]{yifan_sce}%
  \BibitemOpen
  \bibfield  {author} {\bibinfo {author} {\bibfnamefont {P.}~\bibnamefont
  {Abratenko}} \emph {et~al.} (\bibinfo {collaboration} {MicroBooNE}),\
  }\href@noop {} {\bibfield  {journal} {\bibinfo  {journal} {JINST}\ }\textbf
  {\bibinfo {volume} {15}},\ \bibinfo {pages} {P07010} (\bibinfo {year}
  {2020}{\natexlab{a}})}\BibitemShut {NoStop}%
\bibitem [{\citenamefont {Abratenko}\ \emph
  {et~al.}(2020{\natexlab{b}})\citenamefont {Abratenko} \emph
  {et~al.}}]{mike_sce}%
  \BibitemOpen
  \bibfield  {author} {\bibinfo {author} {\bibfnamefont {P.}~\bibnamefont
  {Abratenko}} \emph {et~al.} (\bibinfo {collaboration} {MicroBooNE}),\
  }\href@noop {} {\bibfield  {journal} {\bibinfo  {journal} {JINST}\ }\textbf
  {\bibinfo {volume} {15}},\ \bibinfo {pages} {P12037} (\bibinfo {year}
  {2020}{\natexlab{b}})}\BibitemShut {NoStop}%
\bibitem [{\citenamefont {Eaton}(1983)}]{alma991015286869705251}%
  \BibitemOpen
  \bibfield  {author} {\bibinfo {author} {\bibfnamefont {M.~L.}\ \bibnamefont
  {Eaton}},\ }\href@noop {} {\emph {\bibinfo {title} {Multivariate statistics :
  a vector space approach}}},\ Wiley series in probability and mathematical
  statistics. Probability and mathematical statistics.\ (\bibinfo  {publisher}
  {Wiley},\ \bibinfo {address} {New York},\ \bibinfo {year} {1983 -
  1983})\BibitemShut {NoStop}%
\bibitem [{\citenamefont {Aguilar-Arevalo}\ \emph
  {et~al.}(2009{\natexlab{c}})\citenamefont {Aguilar-Arevalo} \emph
  {et~al.}}]{mb}%
  \BibitemOpen
  \bibfield  {author} {\bibinfo {author} {\bibfnamefont {A.}~\bibnamefont
  {Aguilar-Arevalo}} \emph {et~al.} (\bibinfo {collaboration} {MiniBooNE}),\
  }\href@noop {} {\bibfield  {journal} {\bibinfo  {journal} {Phys. Rev. Lett.}\
  }\textbf {\bibinfo {volume} {102}},\ \bibinfo {pages} {101802} (\bibinfo
  {year} {2009}{\natexlab{c}})}\BibitemShut {NoStop}%
\bibitem [{\citenamefont {Bertuzzo}\ \emph {et~al.}(2018)\citenamefont
  {Bertuzzo} \emph {et~al.}}]{Bertuzzo:2018itn}%
  \BibitemOpen
  \bibfield  {author} {\bibinfo {author} {\bibfnamefont {E.}~\bibnamefont
  {Bertuzzo}} \emph {et~al.},\ }\href {\doibase 10.1103/PhysRevLett.121.241801}
  {\bibfield  {journal} {\bibinfo  {journal} {Phys. Rev. Lett.}\ }\textbf
  {\bibinfo {volume} {121}},\ \bibinfo {pages} {241801} (\bibinfo {year}
  {2018})}\BibitemShut {NoStop}%
\bibitem [{\citenamefont {Ballett}\ \emph {et~al.}(2019)\citenamefont {Ballett}
  \emph {et~al.}}]{Ballett:2018ynz}%
  \BibitemOpen
  \bibfield  {author} {\bibinfo {author} {\bibfnamefont {P.}~\bibnamefont
  {Ballett}} \emph {et~al.},\ }\href {\doibase 10.1103/PhysRevD.99.071701}
  {\bibfield  {journal} {\bibinfo  {journal} {Phys. Rev. D}\ }\textbf {\bibinfo
  {volume} {99}},\ \bibinfo {pages} {071701} (\bibinfo {year}
  {2019})}\BibitemShut {NoStop}%
\bibitem [{\citenamefont {Abdallah}\ \emph {et~al.}(2020)\citenamefont
  {Abdallah}, \citenamefont {Gandhi},\ and\ \citenamefont
  {Roy}}]{Abdallah:2020vgg}%
  \BibitemOpen
  \bibfield  {author} {\bibinfo {author} {\bibfnamefont {W.}~\bibnamefont
  {Abdallah}}, \bibinfo {author} {\bibfnamefont {R.}~\bibnamefont {Gandhi}}, \
  and\ \bibinfo {author} {\bibfnamefont {S.}~\bibnamefont {Roy}},\ }\href@noop
  {} {\bibfield  {journal} {\bibinfo  {journal} {Preprint}\ } (\bibinfo {year}
  {2020})},\ \Eprint {http://arxiv.org/abs/2010.06159} {arXiv:2010.06159
  [hep-ph]} \BibitemShut {NoStop}%
\end{thebibliography}%

\newpage

\end{document}